\documentclass[a4paper]{article}

\usepackage[english]{babel}
\usepackage[utf8x]{inputenc}
\usepackage[T1]{fontenc}

\usepackage[a4paper,top=3cm,bottom=2cm,left=3cm,right=3cm,marginparwidth=1.75cm]{geometry}

\usepackage{amsmath}
\usepackage{graphicx}
\usepackage[colorinlistoftodos]{todonotes}
\usepackage[colorlinks=true, allcolors=blue]{hyperref}
\usepackage{authblk}
\title{\line(1,0){440}\\Investigating the perturbed geometries of vortex rings in free fall through another liquid\\\line(1,0){440}}
\author{Dripto Biswas}
\affil{School of Physical Sciences,

 National Institute of Science Education and Research }
\begin{document}
\maketitle
\begin{abstract}
The goal of this paper is to theoretically investigate the origin of the standing wave-like perturbation observed on the vortex rings falling through another liquid. We simplified the Navier-Stokes equation based on observational evidence from related research, and showed that there exists a steady-state velocity field with the required wave-like characteristics. For better visualization we also plotted the velocity field.
\end{abstract}
\section{Introduction:}
Some liquids are known to form vortex rings when dropped into other liquids having suitable properties \cite{thoms,baumann}.
For example they are seen to form when drops of ink are released into still water(see Fig. \ref{fig_1}). The breakup of the ring into self-similar structures \cite{thoms} is preceded by a wave-like perturbation which is observed to form throughout the vortex ring at almost the same time. The top and side views of these perturbed rings are shown below,
\begin{center}
\fbox{\includegraphics[scale=0.34]{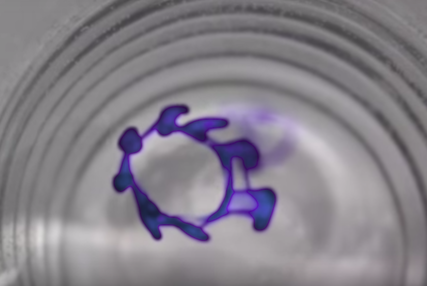} }

\label{fig_1}
Fig. 1: Top View
\end{center}

\begin{center}
\fbox{\includegraphics[scale=0.4]{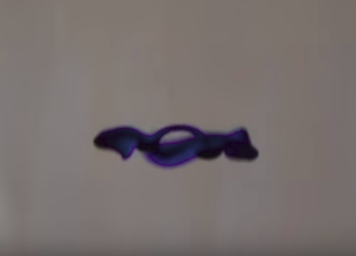}} 

\label{fig_2}
Fig. 2: Side View
\end{center}
The drop to vortex ring formation was first observed and studied by J.J Thomson and H.F Newall  \cite{thoms}.They investigated the dependence of the formation of vortex rings from drops, on various factors like interfacial surface tension between the two liquids, the ratio of $\frac{\eta}{d}$ (which is a measure of the Reynolds number for the falling vortex ring) and initial impact speed  \cite{thoms}. For our purposes, we assume the drop to have been released just below the surface of the water with very low initial velocity. High velocity tends to introduce high circulation in the initial drop, which leads to the breakup of the ring/drop structure(it may diffuse before forming a well-defined toroidal shape) \cite{thoms,baumann,chapman}.

According to the paper by Baumann N. \textit{et al.} \cite{baumann}, low Reynolds number is a characteristic of the flow of the falling vortex ring. We shall use this crucial observation while simplifying the Navier-Stokes Equation.

We assume that the vortex ring has a well-defined toroidal structure (since there is negligible diffusion of the vortex ring \cite{thoms}) and thereby attempt to mathematically model the velocity field of the vortex ring. We apply the Navier-Stokes equation to the volume enclosed by the toroidal surface. Two dimensional plots of the final velocity field, and a three dimensional plot of the perturbation velocity field is provided. The perturbation velocity field is possibly the reason behind the observed standing wave-like pattern on the ring. In the following section, we simplify and solve the Navier-Stokes equation to get the solution representing the falling vortex ring after reaching steady-state.
\section{Mathematical Modelling:}

Treating the ring as a compact object falling through a viscous liquid, we conclude that it must reach a terminal falling velocity, after steady-state flow has been established.
We use a cylindrical coordinate system $(\rho,\phi,z)$, with the origin of coordinates being the geometric centre of the unperturbed toroid, co-moving with the falling vortex ring. 
Let the vortex ring velocity field after terminal velocity has been attained, be given by $\vec \Psi(\rho,\phi,z)$ and the direction of motion(downward) be along the $z$-axis. Note that, we neglect the energy dissipation once steady-state has been attained.

Let $\vec \epsilon = \epsilon(\rho,\phi,z) \hat{k}$ denote the velocity perturbation introduced into the vortex ring along the $z$-axis. The reason behind this directional bias is the acceleration due to gravity vector $\vec g$ pointing along the $z$-axis.

The Navier-Stokes equation is given as  \cite{dobek},
\[\frac{\partial \vec \Psi}{\partial t} + (\vec \Psi \cdot \nabla)\vec \Psi - \nu \nabla^2 \vec \Psi = -\nabla w + \vec g\]
where $w = \frac{p}{d}$, $p$ being the pressure and $d$ the density of the fluid and $\nu$ denotes the kinematic viscosity.

We neglect the $(\vec \Psi \cdot \nabla)\vec \Psi$ (convective) term as the Reynolds number for the flow was observed to be quite less  \cite{baumann}, implying the dominance of viscous forces over inertial forces. Therefore,
\begin{equation}
\frac{\partial \vec \Psi}{\partial t} - \nu \nabla^2 \vec \Psi = -\nabla w + \vec g 
\label{eqn_1}
\end{equation}

We consider $\nu$ to be a constant, as all related parameters(like temperature, density of liquid etc.) are constant. Let $\vec \Phi = \vec \Psi + \vec \epsilon$ denote the perturbed velocity field. 
Since $\vec \Phi$ also must satisfy the Stokes equation, we have,
\[\frac{\partial \vec \Phi}{\partial t} - \nu \nabla^2 \vec \Phi = -\nabla (w+w') + \vec g \]

or, 
\begin{equation}
\frac{\partial \vec \Psi}{\partial t} - \nu \nabla^2 \vec \Psi + \frac{\partial \vec \epsilon}{\partial t} - \nu \nabla^2 \vec \epsilon = -\nabla w -\nabla w' + \vec g . \label{eqn_2}
\end{equation}

Using eqn. \ref{eqn_1} for $\vec \Psi$ and eqn. \ref{eqn_2}, we now have,
\begin{equation}
\frac{\partial \vec \epsilon}{\partial t} - \nu \nabla^2 \vec \epsilon = -\nabla w'.
\label{eqn_3}
\end{equation}

As discussed above, the vortex ring is assumed to have attained a terminal falling velocity.
We are looking for stationery state solutions to the Navier-Stokes equation, since any system under the influence of balanced external forces tend to evolve into stationery states. In other words,
\begin{equation}
\frac{\partial \vec\Phi}{\partial t} = \frac{\partial \vec \Psi}{\partial t} + \frac{\partial \vec \epsilon}{\partial t}=0.
\label{eqn_4}
\end{equation}

But, with respect to the co-moving origin, the original flow field $\vec \Psi$ is time-independent once terminal velocity is achieved(since the energy dissipation is neglected after the ring reaches terminal velocity)
This implies that,
\[\frac{\partial \vec \Psi}{\partial t} = 0 \]
Thus, from eqn. \ref{eqn_4} , we observe that $\vec \epsilon(\rho,\phi,z)$ is a stationery time-independent perturbation. It follows that 
\begin{equation}
\frac{\partial \vec \epsilon}{\partial t} = 0.
\label{eqn_5}
\end{equation} 
Also, if $\nabla w' \ne 0$,
there exists a net external force, which in turn will make the velocity field non-stationary(time-dependant). Therefore, $\nabla w' = 0$. Eqn. \ref{eqn_3} then simplifies to,
\begin{equation}
\nabla^2 \vec \epsilon = 0.
\label{eqn_6}
\end{equation}

Now, we observe that $\epsilon(\rho,\phi,z) \hat{k}$, has no radial or angular component but only a component along the $z$-axis. The vector Laplacian in cylindrical coordinates is given by  \cite{nasser},
\[\nabla^2 \vec A = (\nabla^2 A_{\rho} - \frac{A_{\rho}}{\rho^2} - \frac{2}{\rho^2}\frac{\partial A_\phi}{\partial \phi})\hat{r} + (\nabla^2 A_\phi - \frac{A_\phi}{\rho^2} + \frac{2}{\rho^2}\frac{\partial A_\phi}{\partial \phi})\hat{\phi} + \nabla^2 A_z \hat{k}\]
Substituting $\vec \epsilon$ for $\vec A$ above, and noting that $\epsilon_{\rho} = \epsilon_\phi = 0$, we further simplify eqn. \ref{eqn_6} as,
\begin{equation}
\nabla^2 \vec \epsilon = \nabla^2 \epsilon \hat{k} = 0.
\label{eqn_7}
\end{equation}
This is the Laplace equation, whose solutions are known as harmonic functions. Given appropriate boundary conditions, we can find the exact three dimensional harmonic solutions on a toroid.
  
We proceed to look for suitable solutions of the Laplace equation on a toroid of inner radius $a$ and outer radius $b$, having boundary velocity profiles which are $2\pi$ periodic in nature. The most general solution to the Laplace equation can be found by assuming the separability of components in the solution as  \cite{nasser},
\begin{equation}
\epsilon(\rho,\phi,z) = R(\rho)G(\phi)Z(z).
\label{eqn_8}
\end{equation}

Solving the equation and separating the components we get,
\[Z(z) = A_z e^{kz}+B_z e^{-kz}\]
\begin{equation}
G(\phi) = A_{\phi}\cos(n\phi) + B_{\phi}\sin(n\phi)
\label{eqn_9}
\end{equation}

where $n$ is a positive integral constant and $k$ is a real number.

Now, we note an important feature of $\epsilon(\rho,\phi,z)$, due to the continuity equation. Since, $\nabla \cdot \vec \Phi=0$ and $\nabla \cdot \vec \Psi =0 $ (the flow being incompressible), we must have $\nabla \cdot \vec \epsilon = 0$. This leads to the equation,
\[\frac{\partial \epsilon}{\partial z} = 0\]

However, from eqn. \ref{eqn_8}  we see that this implies $k=0$ in the expression for $Z(z)$. In other words, the perturbation $\vec \epsilon = \epsilon(\rho,\phi) \hat{k}$ is independent of $z$.
The remaining differential equation for the $R(\rho)$ term is given by,
\[\frac{\rho ^2}{R}\frac{d^2 R}{d\rho^2} + \frac{\rho}{R}\frac{dR}{d\rho} = n^2\] whose general solution is,
\begin{equation}
R(\rho) = A_{\rho} \cosh(n \log \rho) + i B_{\rho} \sinh(n \log \rho)
\label{eqn_10}
\end{equation}
where $i = \sqrt{-1}$. We consider only the real part, and write the final solution as,
\[\epsilon = C(a,b) \cosh(n \log \rho) \sin(n\phi)\]
where $C(a,b)$ is a combination of the constants, and it depends on at least 2 initial conditions, most possibly the inner and outer radius $a$ and $b$. $n$ denotes the number of nodes observed on the vortex ring, which we suspect depends on the total energy of the system.
Obviously, this is not the general solution, but we shall use this form for further plots.

\section{Results:}
The boundary conditions can theoretically be determined by careful observation using precise instruments, and is currently beyond the scope of this paper. We have simply chosen a set of values for the constants(based on trial and error) and have constructed a plot of the solution on a toroid of inner radius $a = 0.5$ $cm$ and outer radius $b=0.6$ $cm$
The following are the three dimensional volume plots of the perturbation velocity field $\vec \epsilon$. It is understood that this field is superposed along the $\hat{k}$ direction on $\vec \Psi$. The positive and negative values at various points of the plot correspond to the velocity magnitudes at those points, having directions $\hat{k}$ and $-\hat{k}$ respectively.
\begin{center}
\fbox{\includegraphics[scale=0.3]{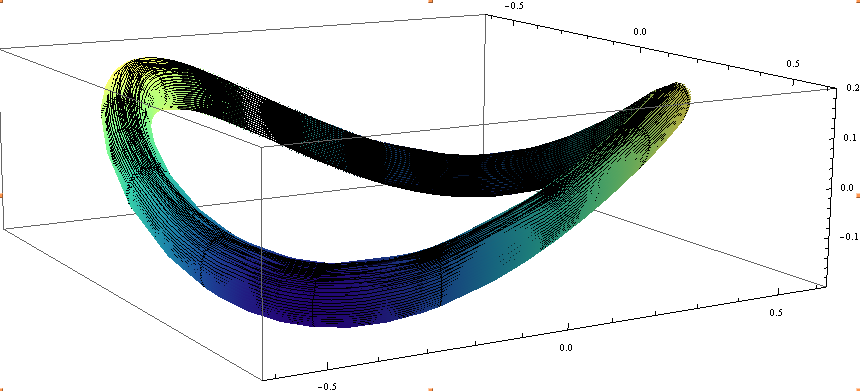}}
\fbox{\includegraphics[scale=0.3]{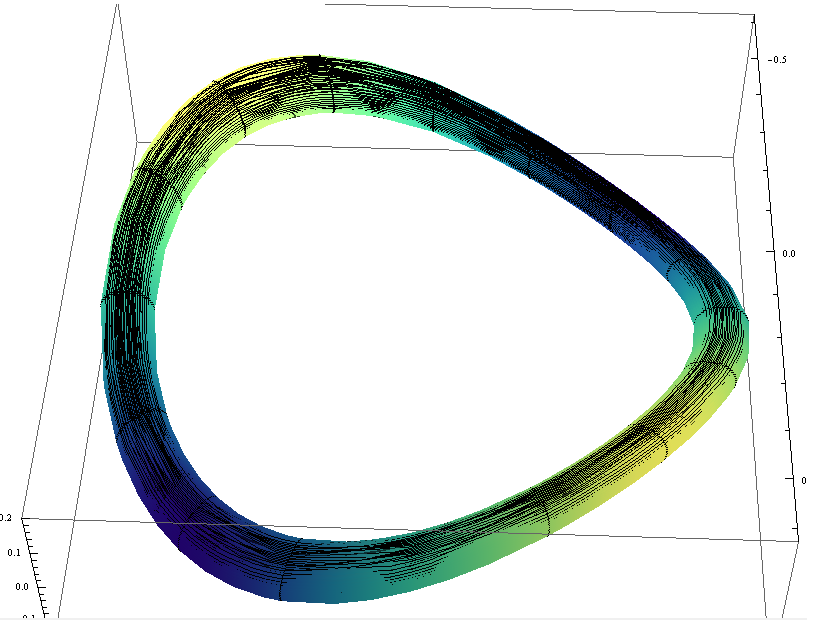}}
\label{fig_3}

Fig. 3:  
3-D plot of $\vec \epsilon = \epsilon \hat{k}$(2 nodes)

Yellow - positive values ; Blue - negative values
\end{center}

\begin{center}
\fbox{\includegraphics[scale=0.4]{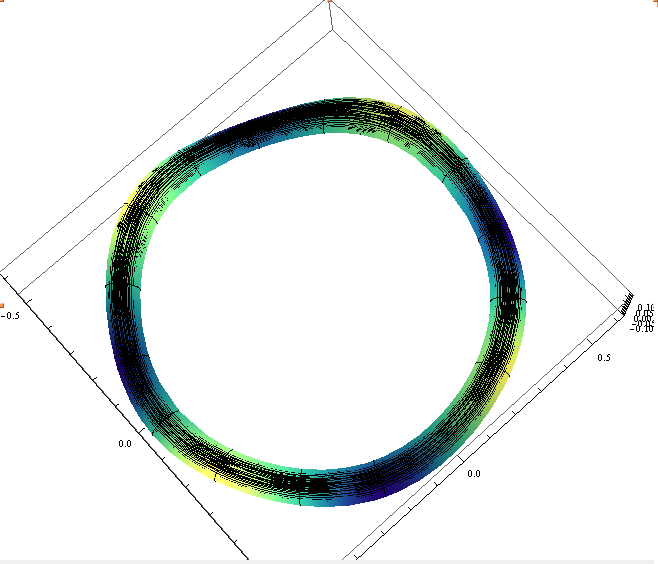}}
\label{fig_4}

Fig. 4: 3-D plot of $\vec \epsilon = \epsilon \hat{k}$(4 nodes)

Yellow - positive values; Blue - negative values
\end{center}

It is not difficult to imagine that the shape of the toroidal ring will be the same (almost) as the one depicted here, as every point inside the toroid develops a new velocity component along $z$-axis, consequently pushing it upwards or downwards depending on $\phi$.

To give an idea of the perturbed velocity field $\vec \Phi$, we consider the circulation field to be represented by a Lamb-Oseen vortex having constant total circulation $\Gamma$ and a core vortex radius of $0.01$ $cm$. Our origin for the subsequent plots, is the centre of the circular cross-section of the toroid, and the two dimensional graphs represent the corresponding cross-section plane of the toroid. We assume that after attaining terminal velocity, the dissipation of energy of the system being negligible, the core radius is preserved. The two dimensional velocity field is then given by,
\begin{equation}
\vec V(r,\theta) = \frac{\Gamma}{2\pi r}(1 - e^{\frac{-r^2}{r_c ^2}}) \hat{\theta}
\label{eqn_11}
\end{equation}

where $r_c = 0.01 cm$ denotes the core radius. Note that this $r$ is different from the $\rho$ in the equations derived above. For the plot, $\rho = 0.55 + r\cos\theta$ from simple geometry. Thus, the perturbation field $\vec \epsilon(\rho,\phi,z)$ can be written in terms of $(r,\theta)$ for given values of $\phi$. We chose to plot the total velocity field at a node ($\phi = \frac{\pi}{4}$) and an anti-node($\phi = \frac{3\pi}{4}$) for a 2-node vortex ring. The plots are shown below.

\begin{center}
\includegraphics[scale=0.6]{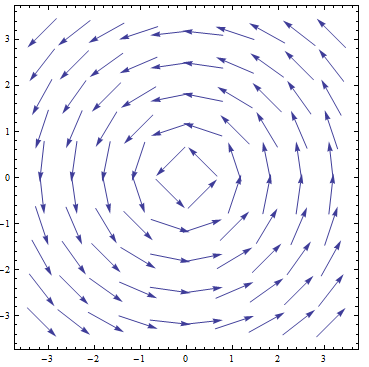} 
\label{fig_5}

Fig. 5: Unperturbed velocity field of a Lamb-Oseen vortex

\includegraphics[scale=0.6]{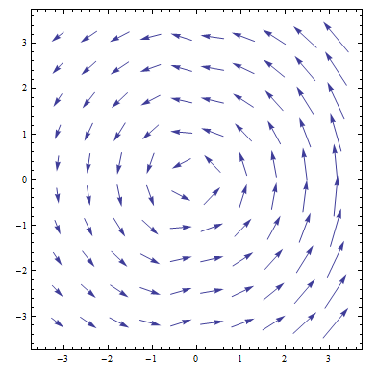} 
\label{fig_6}

Fig. 6: Perturbed velocity field at $\phi = \frac{\pi}{4}$
\end{center}

\begin{center}
\includegraphics[scale=0.6]{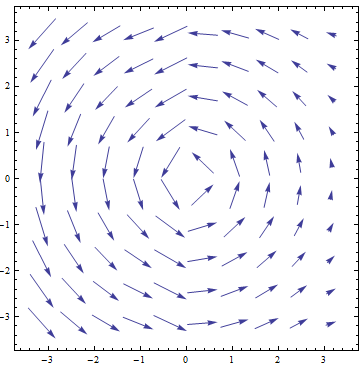} 
\label{fig_7}

Fig. 7: Perturbed velocity field at $\phi = \frac{3\pi}{4}$
\end{center}

Note how, in Fig. 6 and Fig. 7, the velocity field on one side is smaller than the other side. This is due to every point being forced upwards/downwards on superposition of $\vec \epsilon$ to $\vec \Psi$. In Fig. 6, the center of mass of the system(the origin in this case) develops a net upward velocity, as there is greater momentum in the upward direction compared to the downward direction.

\section{Conclusion:}
We developed a theoretical model of the perturbed vortex ring, in an attempt to justify the symmetric wave-like patterns observed on it. The angular component, is clearly responsible for the standing wave-like nature of the perturbation. 

Experiments may be performed to show that these rings indeed reach a terminal velocity before the observed perturbations set in. For our approximate model to be correct, it is essential for the system to reach a steady-state while falling through the liquid.

We conclude with one last remark on what might happen after the perturbation sets in. This should primarily depend on the total circulation $\Gamma$ of the vortex ring. Note that the for the perturbation field $\vec \epsilon$ has a vorticity given by $\omega = \nabla \times \vec \epsilon$. Also, the total circulation due to $\omega$ can be calculated as $\Gamma_{\epsilon} = \int \int_S \omega \cdot ds$, where the integral is over the bounding surface of the toroid. Now, if $\frac{\Gamma_{\epsilon}}{\Gamma} << 1$, the perturbation creates a standing wave-like pattern with low amplitude, but should not disintegrate the ring, as the circulation still remains strong enough to maintain a compact structure. In the two dimensional plots shown above, this will lead to all the 3 plots looking similar to each other with slight differences. However, if $\frac{\Gamma_{\epsilon}}{\Gamma}$ is close to 1, soon after forming the perturbation, the ring should disintegrate, possibly into new drops(branching off from the anti-nodes). In the two dimensional plots, we would observe one side of the circular cross-section of the ring to move much faster compared to the other side, thereby de-stabilising the circulation of the vortex ring. Even this claim may be verified with further experiments which can study the relation between unstable or stable vortex rings and the ratio $\frac{\Gamma_{\epsilon}}{\Gamma}$. Qualitatively, the circulation due to $\vec \epsilon$ should be much smaller than the circulation due to the unperturbed velocity field of the vortex ring, in order to maintain a well-defined toroidal shape.

\section{Acknowledgement:}
I would like to thank Professor Victor Roy of the School of Physical Sciences, NISER , without whose guidance, this theoretical investigation would not have been possible. I would also like to thank the academic administration of NISER for providing me the opportunity to undertake this interesting project.


\begin{thebibliography}{9}
\bibitem{thoms} Thomson J.J, Newall H.F. 1885. On the formation of vortex rings by drops falling into liquids, and some allied phenomena. Proc. R. Soc. Lond. 1886 39, 417-436.
\bibitem{nasser} Prof Dr. I. Nasser, Phys 571,T-131,22-Oct-13
\bibitem{baumann} Baumann N,Joseph DD,Mohr P,Renardy Y.1992. Vortex rings of one fluid in another in free fall.
\bibitem{dobek} Dobek S.2012.Fluid dynamics and the Navier-Stokes Equation.
\bibitem{chapman} Chapman D.S. and Critchlow P.R. 1967.Formation of vortex rings from falling drops. J. Fluid Mech. 29, 177.
\end{thebibliography}
\end{document}